\documentclass[%
 aip,
 amsmath,amssymb,
 reprint,%
]{revtex4-1}

\usepackage{graphicx}
\usepackage{dcolumn}
\usepackage{bm}
\usepackage[colorinlistoftodos]{todonotes}

\usepackage{upgreek}
\usepackage{booktabs}
\usepackage[utf8]{inputenc}
\usepackage[T1]{fontenc}
\usepackage{mathptmx}
\usepackage{etoolbox}
\usepackage[colorlinks=true,linkcolor=blue,citecolor=blue,urlcolor=blue]{hyperref}
\usepackage{siunitx}

\makeatletter
\def\@email#1#2{%
 \endgroup
 \patchcmd{\titleblock@produce}
  {\frontmatter@RRAPformat}
  {\frontmatter@RRAPformat{\produce@RRAP{*#1\href{mailto:#2}{#2}}}\frontmatter@RRAPformat}
  {}{}
}%
\makeatother
\begin{document}

\preprint{AIP/123-QED}

\title[High-Q cryogenic surface acoustic wave resonators in the GHz range]{High-Q cryogenic surface acoustic wave resonators in the GHz range}

\author{A. Tarascio}%
\thanks{These authors contributed equally to this work.}

\author{O. Wicki}
\thanks{These authors contributed equally to this work.}

\author{D. M. Zumbühl}
\affiliation{Department of Physics, University of Basel, Klingelbergstrasse 82, CH-4056 Basel, Switzerland}%
\email{dominik.zumbuhl@unibas.ch}
\date{\today}

\begin{abstract}
Surface acoustic wave (SAW) resonators provide a compact platform for confining microwave-frequency phonons and are widely used in radio-frequency technologies, but their operation at gigahertz frequencies and cryogenic temperatures remains challenging. In this regime, conventional design rules do not directly apply, and achieving high-quality acoustic confinement requires careful consideration about geometry and loss mechanisms. Here, we present a systematic experimental study of SAW resonators on gallium arsenide, a platform of particular interest for hybrid quantum devices but comparatively unexplored for high-$Q$ SAW cavities. By varying key design parameters such as cavity length, wavelength, and crystal orientation, we study resonator performance and achieve quality factors up to $28’000$ in the gigahertz range. In addition, we introduce mesa steps within the acoustic cavity, mimicking realistic device architectures and providing insight into scattering processes and additional dissipation channels. Our results establish practical design guidelines for GaAs-based SAW resonators and support their development as a scalable platform for quantum acoustics and phonon-mediated hybrid systems.
\end{abstract}
\maketitle
\section{Introduction}

Surface acoustic wave (SAW) resonators are compact phononic devices that confine microwave-frequency excitations within sub-millimeter footprints and are widely used used in radio-frequency electronics as filters, oscillators, and delay lines~\cite{Morgan2010}. Beyond these established applications, SAW platforms have recently attracted increasing interest in sensing~\cite{Wu2024,Aslam2023}, optomechanics~\cite{Noguchi2020,Iyer2024}, and hybrid quantum systems~\cite{Schuetz2015,Chu2020,Satzinger2018,Scigliuzzo2025}. Despite the maturity of SAW technology in classical signal processing, its extension to the regime relevant for quantum technologies remains challenging: operation at gigahertz frequencies and cryogenic temperatures modifies the dominant loss mechanisms such that established design heuristics do not directly apply~\cite{Kandel2024,Manenti2016}.

In the context of quantum technologies, SAWs have been implemented in both traveling-wave and resonant configurations~\cite{Krenner2026}. Traveling-wave devices enable functionalities such as delay lines~\cite{Gustafsson2014} and the controlled transport of single electrons via moving acoustic potentials~\cite{Naber2006,Chen2015,Takada2019,Shaju2025,Bertrand2016}. By contrast, resonant structures confine phonons within a cavity, enabling enhanced interactions and longer storage times~\cite{Noguchi2017,Satzinger2018}, but require precise control over geometry and loss mechanisms to achieve high quality factors.

A variety of material platforms have been explored, from strongly piezoelectric lithium niobate~\cite{Bienfait2019,Gruenke2024,Kandel2024,Sasaki2026,Luschmann2023} and low-loss quartz~\cite{Manenti2016,Manenti2017} to thin-film piezoelectrics such as aluminum nitride~\cite{Jiang2023}.
Gallium arsenide (GaAs), by contrast, remains comparatively underexplored despite its unique advantages, including the ability to host high-mobility 2D electron gases~\cite{Manfra2014}, spin qubits~\cite{Petta2005}, single-photon emitters~\cite{Weiss2018}, edge magnetoplasmons~\cite{Bosco2019,Tarascio2026}, and high-impedance microwave resonators~\cite{Scarlino2022,Oppliger2026,Scigliuzzo2025}. SAW-induced effects on electronic transport have been demonstrated in GaAs-based quantum dot systems~\cite{Ebbecke2005,Zhu2025}. However, the integration of SAWs into resonant architectures analogous to cavity quantum electrodynamics (cQED)~\cite{Schuetz2015}, where discrete acoustic modes coherently couple to quantum degrees of freedom, remains largely unexplored.

Here, we present a systematic study of SAW resonators on GaAs at cryogenic temperatures, focusing on the interplay between geometry and acoustic confinement. We investigate how key design parameters such as cavity length, wavelength, and crystal orientation affect the resonator quality factor, achieving internal quality factors up to $28’000$ in the gigahertz regime. In addition, we explore the impact of controlled etched steps within the cavity, providing insight into scattering and additional loss channels for more complex designs that will be required in future quantum devices~\cite{Krenner2026}.
Our results support the development of SAW resonators as a platform for quantum acoustics, where confined phonons can be treated in analogy with photons in cavity quantum electrodynamics~\cite{Luepke2024,Schuetz2015,Manenti2017}.

\section{Experimental results}

\subsection{Device Geometry}

\begin{figure}[h!]
\includegraphics[scale=1]{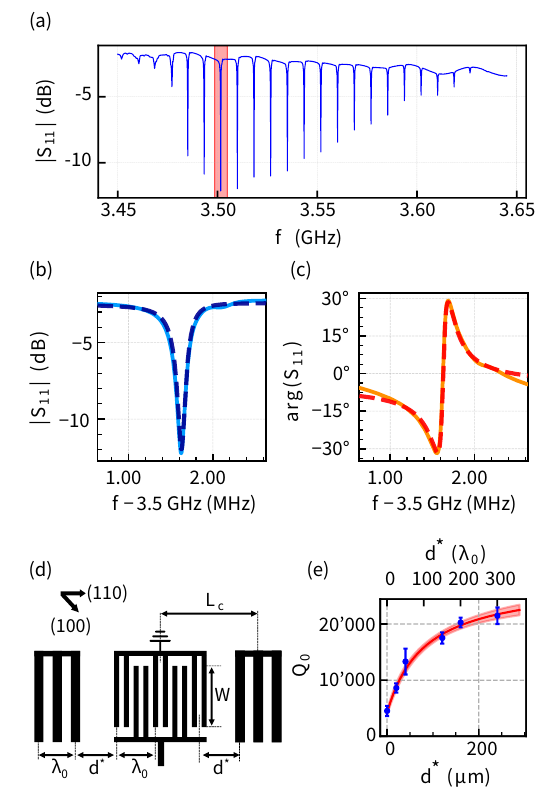}
\caption{\label{Fig_1}(a) Reflection spectrum of a SAW resonator with a cavity $d^*=200.5~{\lambda_0}$. (b) and (c) magnitude and phase of the resonant feature highlighted in panel (a). The dashed lines show the complex fit used to extract the quality factor. (d) Schematics of a one-port SAW resonator using the double-electrode design for the IDT. The resonator is aligned with the [110] crystal direction. (e) Internal quality factor $Q_0$ as a function of cavity size $d^*$. The error bars are obtained from measurements on several devices and the red band shows a fit to Eq.~\ref{PropLoss}. } 
\end{figure}

The typical spectrum of a one-port SAW resonator is shown in Fig.~\ref{Fig_1}a, where multiple longitudinal modes appear within a stop band. The device geometry is illustrated in Fig.~\ref{Fig_1}b. It consists of an interdigitated transducer (IDT) embedded between two distributed Bragg reflectors (DBRs), forming an acoustic cavity. An RF voltage applied to the IDT excites SAWs via the piezoelectric effect, while the DBRs act as frequency-selective mirrors that confine the waves, analogous to a Fabry--Pérot cavity.
The resonant frequency is given by $f_0 = \mathrm{v} / \lambda_0$, where $\lambda_0$ is the acoustic wavelength and $\mathrm{v}$ is the SAW velocity.

The distributed Bragg reflectors (DBRs) consist of shorted electrodes with pitch $\lambda_0/2$, forming a stop band around the design frequency $f_0$. Their reflectivity is set by the number of reflector electrodes $N_{\rm DBR}$ and by the single-electrode reflectivity $r_s$, which depends on the acoustic contrast between free and metallized regions~\cite{Morgan2010,Takasu2019,Kandel2024}. From the measured mirror response we extract $r_s \simeq 3.4\%$ (see Supplementary Information), which for $N_{\rm DBR}=500$ gives near-unity reflectivity and a stop-band width $\Delta f_{\rm DBR} \approx 2 f_0 |r_s|/\pi$.

In the one-port geometry, the IDT both excites and detects the acoustic modes. Its finite length gives a transduction bandwidth of order $\Delta f_{\rm IDT} \approx 1.8 f_0/N_{\rm IDT}$, which, together with the DBR stop band, shapes the spectrum in Fig.~\ref{Fig_1}a. We use a double-electrode IDT to suppress internal Bragg reflection from the transducer itself; a single-electrode IDT with the same periodicity would act as a weak reflector and perturb the cavity modes~\cite{Morgan2010}. A residual velocity mismatch between metallized and free regions nevertheless remains, resulting in acoustic scattering at the IDT and effectively splitting the resonator into two  sections. Each DBR defines an acoustic cavity on either side of the IDT. In contrast to the conventional definition of the cavity length $d$ as the distance between the two DBRs, we define $d^*$ as the distance between the IDT and a single DBR. By including the finite penetration of the SAW into the gratings, the effective cavity length becomes $L_c = d^* + \lambda_0/(2|r_s|)$, where the second term accounts for the penetration length into the two reflectors. Multiple longitudinal modes are therefore supported within each cavity, with free spectral range
\begin{equation}
    \Delta f_{\rm FSR} =
    \frac{f_0}{2 d^*/\lambda_0 + 1/|r_s|}.
    \label{eq:fsr}
\end{equation}
This relation shows that the mode spacing can be engineered through the cavity length $d^*$, allowing the resonator spectrum to be adjusted to address systems operating at different frequencies.

Devices are fabricated using \qty{3}{\nano\meter} Ti / \qty{40}{\nano\meter} Al electrodes. Gold electrodes yielded no clear resonances, consistent with degraded performance reported in the literature~\cite{Takasu2019}.  Our reference design uses a cavity length $d^* = 200.5~\lambda_0$, together with $N_{\mathrm{IDT}} = 39$, $N_{\mathrm{DBR}} = 500$, aperture $W = 45~\mu$m, and $\lambda_0 = 800$ nm. This geometry serves as a baseline for the variations explored in this work. The half-integer values of $d^*$ arise from positioning the DBRs according to the quasi-constant acoustic reflection periodicity (QARP) design~\cite{Ebata1988,Kandel2024}, in which the reflectors are implemented as a direct extension of the IDT geometry, preserving the same periodicity while the non-metallized regions are defined as empty gaps in the electrode pattern.

\subsection{Cavity size}

The internal quality factor of a SAW resonator is determined by a competition between losses at the reflectors and propagation losses within the cavity. The contribution associated with the finite reflectivity of the DBRs gratings can be written as~\cite{Morgan2010}
\begin{equation}
    Q_g = \frac{\omega_0 L_c}{2\rm{v}(1-|\Gamma|)},
\end{equation}
where $\Gamma = \tanh(N_{\rm DBR} r_s)$ is the total reflectivity of the DBRs.

By measuring resonators with a range of cavity sizes $d^*$, as shown in Fig.~\ref{Fig_1}(c), we observe an initial linear increase of $Q_0$ with $d^*$, followed by a saturation for $d^* \gtrsim 100~\lambda_0$. While the linear regime is well captured by $Q_g \propto L_c$, this saturation indicates the onset of additional loss mechanisms.

To account for this behavior, propagation losses characterized by an attenuation coefficient $\alpha_p$ must be included. The resulting internal quality factor can be expressed as~\cite{Manenti2016}
\begin{equation}\label{PropLoss}
    Q_0 = \left( \frac{1}{Q_g} + \frac{\rm v \alpha_p}{\pi f_0} \right)^{-1},
\end{equation}
which captures the crossover from a reflector-limited regime at small $d^*$ to a propagation-limited regime at larger cavity sizes.

To quantify this behavior, we fabricated devices with cavity sizes ranging from $0.5$ to $300.5~\lambda_0$, with $\lambda_0 = 800$~nm, while keeping all other parameters fixed. For each device, the resonance exhibiting the highest $Q_0$ was selected, and when multiple nearby modes were present, the extracted values were averaged. Devices with identical geometries were further averaged to obtain the data shown in Fig.~\ref{Fig_1}(c).

Fitting Eq.~\ref{PropLoss} to the data yields a propagation loss coefficient $\alpha_p = 0.14 \pm 0.01~\mathrm{mm^{-1}}$, associated with a phonon mean free path $l = 1/\alpha_p \approx \qty{7}{\milli\meter}$~\cite{Manenti2016}. In the fit, we assume a fixed penetration depth $L_p = \lambda_0/(4 r_s) \approx 7~\lambda_0$, using $r_s = 3.4\%$, independently extracted from the dependence of the mode spacing $\Delta f_{\rm FSR}$ on $d^*$ (see Supplementary Material).

\subsection{Frequency dependence}

We next investigate the dependence of the internal quality factor $Q_0$ on resonance frequency by scaling the device dimensions, having $\lambda_0$ from $575$ to $1200$~nm, corresponding to frequencies of approximately $2.4$--$4.8$~GHz. All other geometric parameters are kept fixed in units of the wavelength.

\begin{figure}[h!]
\includegraphics[width=8.5cm]{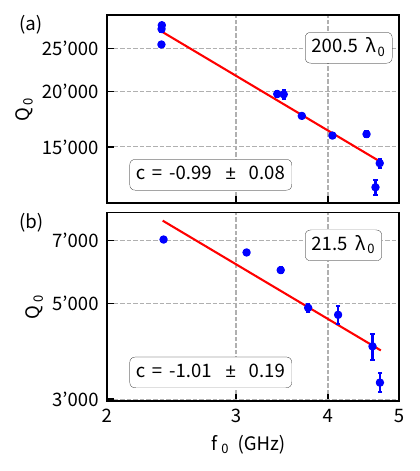}
\caption{\label{Fig_2} Internal quality factor $Q_0$ as a function of resonance frequency $f_0$ for devices with cavity sizes of (a) $200.5~\lambda_0$ and (b) $21.5~\lambda_0$. Error bars indicate the standard deviation of $Q_0$ across different resonances of the same device. Red lines show fits to $Q_0 \propto f_0^{c}$ over the measured frequency range, with the extracted exponents indicated.}
\end{figure}

Two sets of devices were fabricated with cavity sizes $d^* = 21.5~\lambda_0$ and $d^*=200.5~\lambda_0$. The extracted $Q_0$ values, averaged over the resonances of each device, are shown in Fig.~\ref{Fig_2}. Over this limited frequency range, the data are consistent with an empirical power-law dependence $Q_0 \propto f_0^{c}$, resulting in exponents close to $c \approx -1$ for both cavity sizes. Within the measured range, the decrease of $Q_0$ with increasing frequency is relatively weak. This contrasts with the stronger $f^{-2}$ dependence reported for SAW resonators on ST-X quartz~\cite{Manenti2016}.

The weak degradation observed here suggests that GaAs SAW resonators can retain high quality factors across several gigahertz, supporting their use in high-frequency quantum acoustic devices.

\subsection{Crystal axis orientation}

We investigate the effect of crystal orientation on the internal quality factor by fabricating SAW resonators rotated by an angle $\theta$ up to $45^\circ$ with respect to the GaAs [110] cleaving axis, which defines $\theta = 0^\circ$. For symmetry reasons, this $45^\circ$ range covers all distinct in-plane propagation directions, since the $[110]$ and $[1\bar{1}0]$ directions are equivalent. Two sets of devices with cavity sizes $d^* = 25.5~\lambda_0$ and $200.5~\lambda_0$ were studied. As seen in Fig.~3, $Q_0$ decreases with increasing $\theta$ for both cavity sizes. This behavior can be attributed to the anisotropic elastic properties of GaAs, which lead to diffraction and beam-steering of the propagating SAW away from the nominal propagation direction~\cite{Maznev2003}.

A diffraction-limited quality factor can be estimated as~\cite{Emser2022}
\begin{equation}\label{Q_diffraction}
    Q_d = \frac{5\pi}{|1+\gamma|} \left(\frac{W}{\lambda_0}\right)^2,
\end{equation}
where $W$ is the acoustic aperture and
\begin{equation}
    \gamma = \frac{d}{d\theta} \left[ \tanh^{-1} \left( \frac{1}{\rm v(\theta)} \frac{d \rm v(\theta)}{d\theta} \right) \right],
\end{equation}
with $\rm v(\theta)$ the angle-dependent SAW velocity. For GaAs along the (110) direction, $\gamma = -0.536$~\cite{Slobodnik1978}. While this model captures the qualitative trend of decreasing $Q_0$ with increasing $\theta$, we find that it does not quantitatively reproduce the measured dependence, suggesting that additional loss mechanisms such as scattering at the reflectors and finite-aperture effects may also contribute.

\begin{figure}[h!]
\includegraphics[width=8cm]{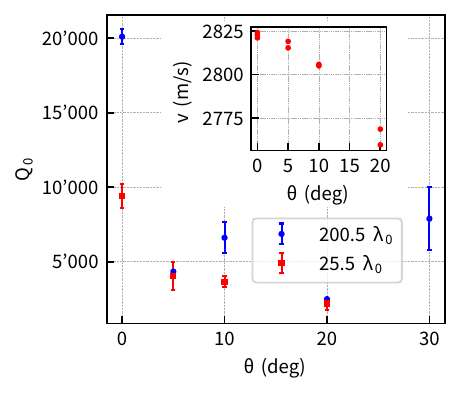}
\caption{\label{Fig_3} Internal quality factor $Q_0$ as a function of in-plane rotation angle $\theta$ for devices with cavity sizes $25.5~\lambda_0$ and $200.5~\lambda_0$. Inset: angle-dependent SAW velocity extracted from the resonance frequency for $d^* = 25.5~\lambda_0$. At $\theta=45^\circ$ the devices show no response.}
\end{figure}

By contrast, the angular dependence of $v$ is consistent with literature reports.~\cite{Powlowski2019}. The velocities extracted from the resonance frequencies, shown in the inset of Fig.~\ref{Fig_3}, follow the expected trend but are systematically lower than reported values~\cite{Powlowski2019,Jungnickel1997,Kuok2001}. This discrepancy is likely due to electrode loading and variations in fabrication. To minimize device-to-device variations, the velocity was extracted from the same resonance across devices when possible, which was not feasible for larger cavities due to the reduced mode spacing.

\subsection{Etched devices}

Surface acoustic waves couple to gate-defined quantum dots and other artificial atoms via the piezoelectric field, enabling interactions between confined charges and phonons~\cite{Schuetz2015,Delsing2019,Krenner2026}. In such devices, mesa steps are typically etched into the heterostructure, introducing loss channels within the acoustic cavity.

\begin{figure}[h!]
\includegraphics[width=8.5cm]{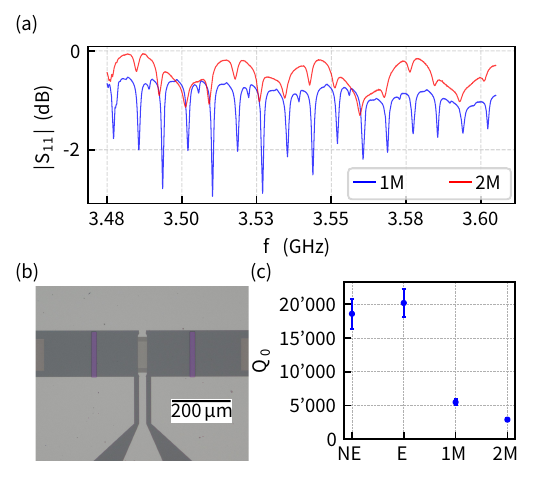}
\caption{\label{fig_5} (a) Reflection spectrum of SAW resonators with cavities with $d^*=200.5~\lambda_0$ with one  mesa placed in one cavity (1M, in blue) and one mesa in each of the cavities (2M in red). (b) Optical microscope picture of a SAW device containing mesas in each cavity, false colored in purple (2M). (c) Internal quality factor for devices fabricated on non-etched (NE), Etched (E) GaAs, devices with a mesa in one cavity (1M) and a mesa in both cavities (2M).}

\end{figure}

To assess their impact, we fabricate resonators with one (1M) or two (2M) mesas inside a $200.5~\lambda_0$ cavity. The steps are \qty{10}{\micro\meter} wide, span the full transverse width of the cavity and are etched down by about \qty{150}{\nano\meter}.  As shown in Fig.~\ref{fig_5}, mesas significantly degrade performance: a single mesa reduces the internal quality factor by about a factor of four, with a further reduction for two mesas. The resonance modes also become weaker and less well defined.

We attribute this behavior to scattering and phase perturbations at the mesa edges. Each mesa introduces step discontinuities that partially reflect Rayleigh waves and convert energy into bulk modes~\cite{Graczykowski2012}, while also modifying the local phase velocity. With two mesas, multiple reflections and additional loss channels further suppress constructive interference, leading to increased dissipation and reduced internal quality factor~\cite{Morgan2010}.

These results demonstrate that while SAW resonators can be integrated with etched device architectures, even sub-wavelength topographic perturbations can significantly impact their performance. Careful design of the mesa geometry and its placement within the cavity is therefore required to minimize additional phase mismatch and scattering losses.

\section{Conclusion}

We have characterized GaAs SAW resonators at cryogenic temperatures in the gigahertz regime, focusing on how geometry controls acoustic confinement. By varying cavity length, wavelength, and propagation direction, we achieve internal quality factors up to $28'000$ and identify practical design guidelines for high-$Q$ operation on GaAs.

The cavity-length dependence shows a crossover from reflector-limited to propagation-loss-limited behavior, corresponding to a phonon mean free path of approximately \qty{7}{\milli\meter}. The quality factor decreases only weakly between $2.4$ and $4.8~\mathrm{GHz}$, while alignment along the GaAs $[110]$ and $[1\bar{1}0]$ directions is favorable for minimizing diffraction-related losses. Finally, we show that etched mesa structures, relevant for integration with gate-defined quantum devices, strongly affect resonator performance. A single mesa reduces $Q_0$ by about a factor of four, indicating that etched steps introduce significant scattering and loss.

Overall, our results establish GaAs as a viable platform for gigahertz SAW resonators compatible with semiconductor quantum-device architectures~\cite{Delsing2019,Krenner2026}. By linking resonator performance to concrete design parameters we open the way for GaAs SAW resonators as a scalable platform for hybrid quantum systems, where confined phonons act as acoustic analogues of cavity QED mediators coupling electronic, optical, and spin degrees of freedom~\cite{Schuetz2015}.
\subsection{Data availability}
The data that support the findings of this study are openly available in the Zenodo repository at http://doi.org/[doi], reference number [reference number].
\subsection{Author Contributions}
A.T. and O.W. designed and fabricated the devices, performed the data analysis and prepared the figures. D.M.Z. supervised the project. A.T. and D.M.Z. wrote the manuscript.

\subsection{Acknowledgments}
We thank Michael Steinacher, Sascha Linder, Sascha Martin, and Sergii Kokhas for their technical support. We are grateful to Pasquale Scarlino for insightful discussions. We also thank Omid Sharifi Sedeh,  Leon C. Camenzind, Janica Böhler, and Simon Svab for valuable discussions, help with the project and fabrication . This research was supported by the Swiss National Science Foundation (grant no. 215757), NCCR SPIN of the SNSF (grant no. 225153), UpQuantVal InterReg. and the Swiss Nanoscience Institute.
The authors declare no conflict of interest.

%

\clearpage
\onecolumngrid
\section*{Supplementary Material}
\setcounter{section}{0}
\renewcommand{\thesection}{S\arabic{section}}
\renewcommand{\thefigure}{S\arabic{figure}}
\renewcommand{\thetable}{S\arabic{table}}
\setcounter{figure}{0}
\setcounter{table}{0}

\subsection{Setup}
The frequency response of the SAW resonators was measured with a vector network analyzer (VNA) from Rohde \& Schwarz at room temperature and approximately 4--5~K. The VNA is connected by SMA cables to a dipstick with low attenuation coaxial cables at which end the PCB containing the devices is connected. The devices are wire bonded with aluminum wires.  At room temperature, the VNA is calibrated to eliminate standing waves in the signal caused by the cables. After the sample is cooled down, the dipstick is slightly retracted before the measurement is made to avoid dissipation introduced by the liquid helium~\cite{Aoki2004} since our system is not under vacuum. To minimize noise, the VNA is set to a power of -3~dBm and bandwidth of 1~KHz.
To ensure consistent convergence of the fit, the data were processed beforehand to eliminate artifacts due to the setup. The standing waves created by the cables cause smooth distortions within the frequency response of the resonator. This smooth background can be removed using an Asymmetric Least Squares Baseline Correction, leaving the resonance dips of the resonator intact.
The finite length of the SMA cables leads to a delay of the signal between the input and output of the VNA. This causes a slope in the phase of the signal. While there are more elaborate procedures to properly account for this delay~\cite{Probst2015}, removing a linear offset of the resonator phase improves the convergence of the resonator fit.

\subsection{Extraction of finger reflectivity, SAW velocity, and mode spacing}
\label{sec:supp_finger_reflectivity_velocity}

The mode spacing provides a direct estimate of the single-finger reflectivity of the DBRs and of the SAW velocity. We model the resonator as an acoustic Fabry--Pérot cavity with an effective length larger than the lithographic cavity size, because the standing wave penetrates into the DBRs. The mode spacing is therefore

\begin{equation}
\Delta f =
\frac{v}{2L_{\mathrm{c}}},
\qquad
L_{\mathrm{c}} = d^* + 2L_p ,
\label{eq:mode_spacing_eff_length}
\end{equation}
where \(d^*\) is the physical separation between the IDT and one DBR and \(L_p\) is the penetration length into one reflector. For a weak periodic reflector, the acoustic amplitude decays inside the grating over a characteristic length

\begin{equation}
L_p \simeq \frac{\lambda_0}{4|r_s|},
\label{eq:penetration_length}
\end{equation}
where \(r_s\) is the single-finger amplitude reflectivity. The factor \(1/|r_s|\) reflects the number of weakly reflecting periods required to build up the mirror response, while the numerical factor follows from the Bragg condition and the phase relation between successive reflections in the grating.

\begin{figure}[h]
\centering
\includegraphics[width=0.85\linewidth]{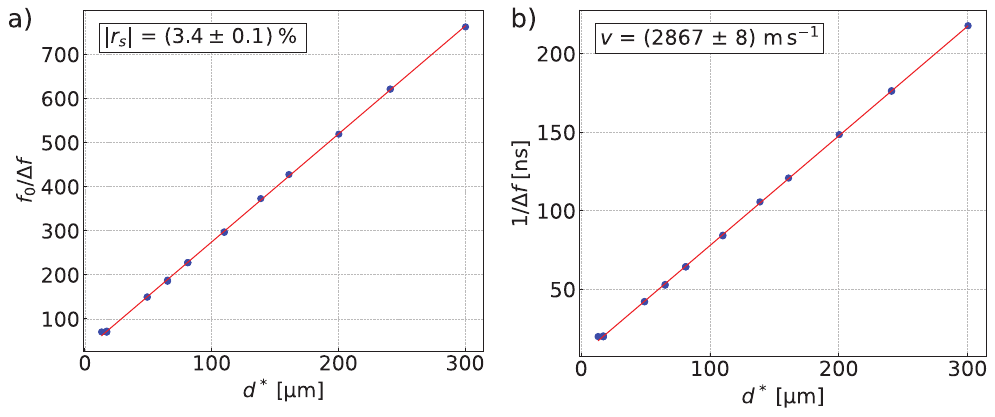}
\caption{
Extraction of the single-finger reflectivity and SAW velocity from the cavity mode spacing.
(a) Inverse normalized mode spacing \(f_0/\Delta f\) as a function of cavity size \(d^*/\lambda_0\). The linear fit follows Eq.~\eqref{eq:fsr}, and the intercept yields the single-finger reflectivity \(|r_s|\).
(b) Inverse mode spacing \(1/\Delta f\) as a function of the physical cavity size \(d^*\). The slope of the linear fit gives the SAW velocity \(v\). Error bars are smaller than the symbol size and are omitted for clarity.
}
\label{fig:reflectivity_velocity}
\end{figure}

Using \(f_0 = v/\lambda_0\), Eqs.~\eqref{eq:mode_spacing_eff_length} and \eqref{eq:penetration_length} give

\begin{equation}
\Delta f =
\frac{f_0}{2d^*/\lambda_0 + 1/|r_s|},
\qquad
\frac{f_0}{\Delta f}
=
\frac{2d^*}{\lambda_0}
+
\frac{1}{|r_s|}.
\label{eq:fsr}
\end{equation}
Thus, a linear fit of \(f_0/\Delta f\) as a function of \(d^*/\lambda_0\) yields the single-finger reflectivity from the intercept. The SAW velocity is extracted independently from a linear fit of \(1/\Delta f\) as a function of the physical cavity size \(d^*\), for which the slope is \(2/v\). The corresponding fits are shown in Fig.~\ref{fig:reflectivity_velocity}.

For the devices studied here, we obtain

\begin{equation}
|r_s| = (3.4 \pm 0.1)\%,
\qquad
v = (2867 \pm 8)\,\mathrm{m\,s^{-1}},
\qquad
L_p \simeq 5.9~\mu\mathrm{m} \simeq 7.4\,\lambda_0 .
\label{eq:extracted_rs_v}
\end{equation}

The extracted value of \(r_s\) also determines the DBR mirror strength. In the weak-reflection limit, the total amplitude reflectivity of a DBR with \(N_{\rm DBR}\) fingers is approximately

\begin{equation}
|\Gamma| \simeq \tanh \left( N_{\rm DBR} |r_s| \right).
\label{eq:dbr_reflectivity}
\end{equation}

For \(N_{\rm DBR}=500\), this gives \(N_{\rm DBR}|r_s| \gg 1\), corresponding to near-unity reflectivity at the center of the stop band. The corresponding DBR stop-band width can be estimated as

\begin{equation}
\Delta f_{\rm DBR}
\simeq
\frac{2 f_0 |r_s|}{\pi}.
\label{eq:dbr_bandwidth}
\end{equation}

This bandwidth sets the frequency range over which the acoustic mirrors provide efficient confinement. Modes far from the stop-band center experience reduced mirror reflectivity and therefore weaker confinement.

In a one-port resonator, the IDT acts both as the emitter and detector of SAWs. Because the IDT has a finite number of electrodes, its transduction efficiency is frequency dependent and follows an envelope set by its spatial Fourier transform. To leading order, the accessible bandwidth scales inversely with the number of IDT electrodes~\cite{Manenti2016,Morgan2010},

\begin{equation}
\Delta f_{\rm IDT}
\approx
1.8 \frac{f_0}{N_{\rm IDT}}.
\label{eq:idt_bandwidth}
\end{equation}

The measured spectrum is therefore shaped by two frequency-dependent filters: the DBR stop band determines the range of frequencies over which modes are confined, while the IDT bandwidth determines how efficiently those modes are excited and detected. This explains why modes away from the design frequency appear with reduced amplitude in Fig.~\ref{Fig_1}a.

\subsection{Quality factor fit}
SAW resonators are commonly modeled using the Butterworth von Dyke equivalent circuit~\cite{Kandel2024} to describe their frequency response. Close to the resonance frequency, the resonator can be described by a simpler series RLC circuit, which allows the reflection coefficient to be written as a function of the internal and external quality factor. Using $\delta f = (f-f_0)/f_0$ with $f_0$ being the resonance frequency, the fitting function to extract the internal $Q_0$ and external $Q_{ext}$ quality factors from the resonance is given by Eq.~\ref{fit_func}. 
\begin{equation}\label{fit_func}
     S_{11}(f)=A\cdot\frac{\frac{Q_{ext}-Q_0}{Q_{ext}}+i2Q_0\delta f}{\frac{Q_{ext}+Q_0}{Q_{ext}}+i2Q_0\delta f} \cdot e^{i\phi}
\end{equation}
Here, $\phi$ accounts for an arbitrary sweep starting phase and $A$ for a global scaling factor. The values of interest are the resonance frequency $f_0$ and $Q_0,~ Q_{ext}$. By choosing a complex-valued $Q_{ext}$, we can account for asymmetric resonance shapes. The fitting was performed using an adapted version of the complex resonator model~\cite{MattNewville2024}. Rather than fitting only the magnitude of the data, this model fits the circular shape of the resonance in the complex plane.

\clearpage
\subsection{Measured devices}
\setlength{\tabcolsep}{6pt} 
\begin{table}[h]
\centering
\label{Table_1a}
\begin{tabular}{lrrrrrrrr}
\toprule
Device & $\lambda_0$ (nm) & $d^* / \lambda_0$ & $\theta$ ($^\circ$) & $f_0$ (GHz)& $Q_0 / 10^3$ & $|Q_\mathrm{ext}| / 10^3$ & $f_0 Q_0 / 10^{13}$ & $\Delta\varphi$ (rad) \\
\hline
\midrule
$V1_{P2}$ & 800 & 0.5 & 0 & 3.53 & 3.88 & 7.67 & 1.37 & 1.08 \\
$V4_{P5}$ & 800 & 0.5 & 0 & 3.54 & 5.16 & 9.52 & 1.83 & 1.20 \\
$NE_C$ & 800 & 25.5 & 0 & 3.48 & 8.50 & 51.82 & 2.96 & 0.37 \\
$SW_C$ & 800 & 25.5 & 0 & 3.53 & 10.03 & 35.63 & 3.54 & 0.71 \\
$NW_G$ & 800 & 25.5 & 0 & 3.51 & 9.76 & 32.06 & 3.42 & 0.75 \\
$SE_J$ & 800 & 50.5 & 0 & 3.49 & 11.49 & 49.86 & 4.01 & 0.49 \\
$SW_B$ & 800 & 50.5 & 0 & 3.50 & 13.31 & 30.34 & 4.66 & 0.93 \\
$NE_B$ & 800 & 50.5 & 0 & 3.50 & 16.07 & 35.59 & 5.63 & 0.99 \\
$SE_E$ & 800 & 150.5 & 0 & 3.52 & 17.51 & 423.03 & 6.17 & 0.09 \\
$SE_I$ & 800 & 200.5 & 0 & 3.49 & 19.69 & 286.82 & 6.87 & 0.14 \\
$SE_C$ & 800 & 200.5 & 0 & 3.52 & 19.33 & 337.41 & 6.80 & 0.13 \\
$NW_C$ & 800 & 200.5 & 0 & 3.50 & 19.93 & 438.95 & 6.98 & 0.09 \\
$NW_I$ & 800 & 200.5 & 0 & 3.47 & 20.68 & 356.20 & 7.18 & 0.12 \\
$NW_P$ & 800 & 300.5 & 0 & 3.47 & 23.47 & 249.38 & 8.13 & 0.19 \\
$SE_P$ & 800 & 300.5 & 0 & 3.50 & 21.40 & 232.43 & 7.48 & 0.19 \\
$NE_i$ & 800 & 25.5 & 5 & 3.47 & 4.69 & 166.97 & 1.63 & 0.07 \\
$SW_I$ & 800 & 25.5 & 5 & 3.52 & 3.38 & 111.36 & 1.19 & 0.07 \\
$NE_J$ & 800 & 200.5 & 5 & 3.48 & 4.34 & 224.12 & 1.51 & 0.04 \\
$NE_K$ & 800 & 25.5 & 10 & 3.48 & 3.92 & 57.45 & 1.37 & 0.16 \\
$SW_K$ & 800 & 25.5 & 10 & 3.49 & 3.39 & 43.45 & 1.18 & 0.18 \\
$NE_l$ & 800 & 200.5 & 10 & 3.47 & 7.34 & 303.94 & 2.55 & 0.05 \\
$SW_L$ & 800 & 200.5 & 10 & 3.45 & 5.87 & 145.70 & 2.03 & 0.10 \\
$SE_S$ & 800 & 25.5 & 20 & 3.41 & 2.47 & 88.68 & 0.84 & 0.06 \\
$NW_S$ & 800 & 25.5 & 20 & 3.42 & 1.90 & 34.38 & 0.65 & 0.12 \\
$NW_R$ & 800 & 200.5 & 20 & 3.40 & 2.49 & 151.37 & 0.85 & 0.04 \\
$TF1_O$ & 800 & 200.5 & 30 & 3.38 & 10.01 & 449.92 & 3.39 & 0.05 \\
$TF1_P$ & 800 & 200.5 & 30 & 3.40 & 5.33 & 393.59 & 1.81 & 0.03 \\
$TF2_O$ & 800 & 200.5 & 30 & 3.38 & 9.16 & 255.79 & 3.10 & 0.08 \\
$TF2_P$ & 800 & 200.5 & 30 & 3.39 & 7.07 & 319.07 & 2.39 & 0.04 \\
$NW_M$ & 575 & 200.5 & 0 & 4.64 & 12.16 & 249.55 & 5.64 & 0.11 \\
$SW_F$ & 575 & 200.5 & 0 & 4.71 & 13.78 & 177.57 & 6.48 & 0.16 \\
$SW_D$ & 600 & 200.5 & 0 & 4.51 & 16.02 & 317.45 & 7.23 & 0.10 \\
$SE_V$ & 680 & 200.5 & 0 & 4.06 & 15.90 & 221.03 & 6.45 & 0.15 \\
$SE_D$ & 750 & 200.5 & 0 & 3.69 & 17.61 & 674.37 & 6.49 & 0.05 \\
$SE_B$ & 820 & 200.5 & 0 & 3.41 & 19.73 & 218.40 & 6.74 & 0.19 \\
$SE_K$ & 1200 & 200.5 & 0 & 2.37 & 27.65 & 454.76 & 6.57 & 0.12 \\
$SE_L$ & 1200 & 200.5 & 0 & 2.38 & 28.20 & 325.38 & 6.70 & 0.18 \\
$NW_K$ & 1200 & 200.5 & 0 & 2.37 & 25.51 & 388.42 & 6.06 & 0.13 \\
$FBC_Q$ & 575 & 21.5 & 0 & 4.68 & 2.87 & 48.71 & 1.34 & 0.14 \\
$FBC_K$ & 600 & 21.5 & 0 & 4.60 & 3.98 & 36.58 & 1.83 & 0.23 \\
$FBC_M$ & 680 & 21.5 & 0 & 4.13 & 4.71 & 40.70 & 1.95 & 0.36 \\
$FBC_C$ & 750 & 21.5 & 0 & 3.76 & 4.89 & 27.29 & 1.84 & 0.43 \\
$FBC_H$ & 820 & 21.5 & 0 & 3.45 & 5.98 & 58.42 & 2.06 & 0.21 \\
$FBC_P$ & 920 & 21.5 & 0 & 3.10 & 6.57 & 27.60 & 2.04 & 0.51 \\
$FBC_L$ & 1200 & 21.5 & 0 & 2.39 & 7.04 & 56.75 & 1.68 & 0.25 \\
\hline \hline
\bottomrule
\end{tabular}

\caption{Main dip fit results, with angle in respect with the cleaving direction $\theta$ and phase jump $\Delta\varphi$ are indicated}
\end{table}

\begin{table}[h]  
    \centering
    \label{Table_2}
\begin{tabular}{lrrrrrrrrlr}
\toprule
Device & $\lambda$ (nm) & $d^* / \lambda$ & $\theta$ ($^\circ$) & $f_0$ (GHz) & $Q_0 / 10^3$ & $|Q_\mathrm{ext}| / 10^3$ & $f_0 Q_0 / 10^{13}$ & $\Delta\varphi$ (rad) & Etched & Mesa \\
\hline
$FT1_C$ & 800 & 200.5 & 0 & 3.54 & 24.18 & 63.13 & 8.55 & 0.83 & No & 0 \\
$FT1_A$ & 800 & 200.5 & 0 & 3.52 & 17.30 & 91.35 & 6.09 & 0.39 & No & 0 \\
$FT1_B$ & 800 & 200.5 & 0 & 3.53 & 13.57 & 99.03 & 4.80 & 0.29 & No & 0 \\
$FT2_A$ & 800 & 200.5 & 0 & 3.51 & 19.51 & 104.76 & 6.85 & 0.38 & No & 0 \\
$FT1_D$ & 800 & 200.5 & 0 & 3.51 & 19.12 & 82.58 & 6.72 & 0.47 & Yes & 0 \\
$FT1_M$ & 800 & 200.5 & 0 & 3.52 & 16.22 & 93.60 & 5.71 & 0.35 & Yes & 0 \\
$FT1_G$ & 800 & 200.5 & 0 & 3.51 & 27.77 & 50.84 & 9.76 & 1.20 & Yes & 0 \\
$FT1_J$ & 800 & 200.5 & 0 & 3.52 & 21.91 & 63.64 & 7.72 & 0.75 & Yes & 0 \\
$FT2_D$ & 800 & 200.5 & 0 & 3.51 & 12.29 & 111.33 & 4.31 & 0.22 & Yes & 0 \\
$FT2_G$ & 800 & 200.5 & 0 & 3.52 & 26.60 & 62.37 & 9.36 & 0.92 & Yes & 0 \\
$FT2_J$ & 800 & 200.5 & 0 & 3.51 & 17.82 & 105.03 & 6.26 & 0.35 & Yes & 0 \\
$FT1_E$ & 800 & 200.5 & 0 & 3.53 & 5.43 & 46.42 & 1.91 & 0.23 & Yes & 1 \\
$FT1_K$ & 800 & 200.5 & 0 & 3.52 & 5.47 & 94.82 & 1.93 & 0.11 & Yes & 1 \\
$FT2_K$ & 800 & 200.5 & 0 & 3.52 & 4.74 & 47.91 & 1.67 & 0.21 & Yes & 1 \\
$FT2_E$ & 800 & 200.5 & 0 & 3.51 & 4.77 & 141.53 & 1.68 & 0.07 & Yes & 1 \\
$FT2_I$ & 800 & 200.5 & 0 & 3.52 & 7.03 & 49.40 & 2.47 & 0.31 & Yes & 1 \\
$FT1_F$ & 800 & 200.5 & 0 & 3.52 & 3.20 & 126.35 & 1.13 & 0.05 & Yes & 2 \\
$FT1_L$ & 800 & 200.5 & 0 & 3.51 & 2.89 & 144.54 & 1.02 & 0.04 & Yes & 2 \\
$FT2_L$ & 800 & 200.5 & 0 & 3.52 & 2.57 & 89.72 & 0.91 & 0.25 & Yes & 2 \\
$FT2_F$ & 800 & 200.5 & 0 & 3.52 & 3.52 & 56.43 & 1.24 & 0.13 & Yes & 2 \\
$FT2_H$ & 800 & 200.5 & 0 & 3.51 & 2.36 & 62.24 & 0.83 & 0.08 & Yes & 2 \\
\bottomrule
\hline \hline
\end{tabular}
\caption{Main dip fit results, with angle in respect with the cleaving direction $\theta$, phase jump $\Delta\varphi$ , if the device is lying on an etched substrate, and the presence of mesa steps indicated}
\end{table}

\end{document}